\documentstyle[amssymb,thmsa]{article}

\input tcilatex
\begin{document}

\title{${\Bbb Z}_{n}^{3}$-graded colored supersymmetry}
\author{Bertrand Le Roy \\
Laboratoire de Gravitation et Cosmologie Relativistes\\
Tour {\bf 22}-12, 4$^{{\rm e}}$ \'{e}tage\\
Boite courrier 142\\
Universit\'{e} Pierre et Marie Curie\\
4 place Jussieu\\
75252 PARIS Cedex 05}
\maketitle

\begin{abstract}
We build generalizations of the Grassmann algebras from a few simple
assumptions which are that they are graded, maximally symmetric and contain
an ordinary Grassmann algebra as a subalgebra. These algebras are graded by $%
{\Bbb Z}_{n}^{3}$ and display surprising properties that indicate their
possible application to the modelization of quark fields. We build the
generalized supersymmetry generators based on these algebras and their
derivation operators. These generators are {\em cubic} roots of the usual
supersymmetry generators.
\end{abstract}

\section{Introduction}

Many attempts\cite{sheunert, rittenberg, lukierski, kleeman} have been made
to replace the ${\Bbb Z}_{2}$ grading group that is usually used in
supersymmetry\cite{wesszumino, corwin} by another abelian group. Most of
these attempts deal with $({\Bbb Z}_{2})^{n}$ grading groups and the
corresponding generalized Lie algebras, that are called Color Lie
Superalgebras. It has been shown by Scheunert\cite{sheunert} that any $%
\varepsilon $-Lie superalgebra, of which the Color Lie Superalgebras are
particular cases, is isomorphic to an ordinary Lie superalgebra, leaving
little room for a generalization of supersymmetry based on an alternative
grading group, at least at the level of the Lie superalgebra. Attempts have
also been made to generalize supersymmetry by replacing the Grassmann
algebra with a paragrassmann algebra, but it is not clear what the algebraic
structures corresponding to Lie superalgebras are in this case, and these
algebras are very unsymmetrical: they need an ordering relation between the
generators\cite{mohammedi, chen}. We build a generalization of the Grassmann
algebras from different assumptions and build the corresponding generalized
supersymmetry generators. The resulting algebraic structure is an $%
\varepsilon $-Lie superalgebra (the Jacobi identity is satisfied), and it
has a subalgebra that is a Lie superalgebra.

\section{Generalized Grassmann algebras}

Any grading group ${\cal G}$, being commutative, can be decomposed into the
product of ${\Bbb Z}$ or ${\Bbb Z}_{p}$ groups 
\[
{\cal G}=\prod_{i=1}^{n}{\Bbb Z}_{p_{i}} 
\]
where $p_{i}$ is $0$ by convention for a ${\Bbb Z}$ group. If an element $A$
of a ${\cal G}$-graded structure is of grade $a$, we will note $%
a=\{a_{i};1\leqslant i\leqslant n\}$ to distinguish the integers
caracterizing the grade of $A$ in each of the ${\Bbb Z}_{p_{i}}$ groups. One
can then define a ${\cal G}$-graded commutation factor, whose general form%
\cite{kleeman, sheunert} is 
\[
\varepsilon (a,b)=\prod_{i=1}^{n}(\pm 1)^{a_{i}b_{i}}\prod_{1\leqslant
i<j\leqslant n}q_{ij}^{a_{i}b_{j}-a_{j}b_{i}} 
\]
where the $q_{ij}$ are $r_{ij}^{{\rm th}}$ roots of unity, $r_{ij}$ being
the greatest common divider of $p_{i}$ and $p_{j}$. One can choose $-1$ in
the first factors only in the case where $p_{i}$ is even.

The assumptions we make about our generalized Grassmann algebra are that:

\begin{itemize}
\item  it is $\varepsilon $-commutative, that is 
\[
AB=\varepsilon (g_{A},g_{B})BA
\]
for all $A$ and $B$ in the algebra, where $g_{A}$ and $g_{B}$ are the grades
of $A$ and $B$.

\item  the ${\Bbb Z}_{p_{i}}$ groups are all equivalent and
undistinguishable.

\item  it contains a fermionic sector (formed with elements that anticommute
with each other) that put on an equal footing the ${\Bbb Z}_{p_{i}}$ groups.
\end{itemize}

Any Grassmann algebra satisfying these assumptions with a grading group
composed of the product of two ${\Bbb Z}_{p_{i}}$ groups is isomorphic to an
ordinary Grassmann algebra. No Grassmann algebra with a grading group
composed of the product of four or more ${\Bbb Z}_{p_{i}}$ groups is able to
put them on an equal footing, unless it is isomorphic to an ordinary
Grassmann algebra. This is because it is impossible to build a symmetric
oriented graph with four points. Here are the four possible oriented graphs
with four points:\FRAME{dtbpF}{7.9408cm}{6.3526cm}{0pt}{}{\Qlb{tetragraph}}{%
graphtet.eps}{\special{language "Scientific Word";type
"GRAPHIC";maintain-aspect-ratio TRUE;display "USEDEF";valid_file "F";width
7.9408cm;height 6.3526cm;depth 0pt;original-width 727.625pt;original-height
581.375pt;cropleft "0";croptop "1";cropright "1";cropbottom "0";filename
'c: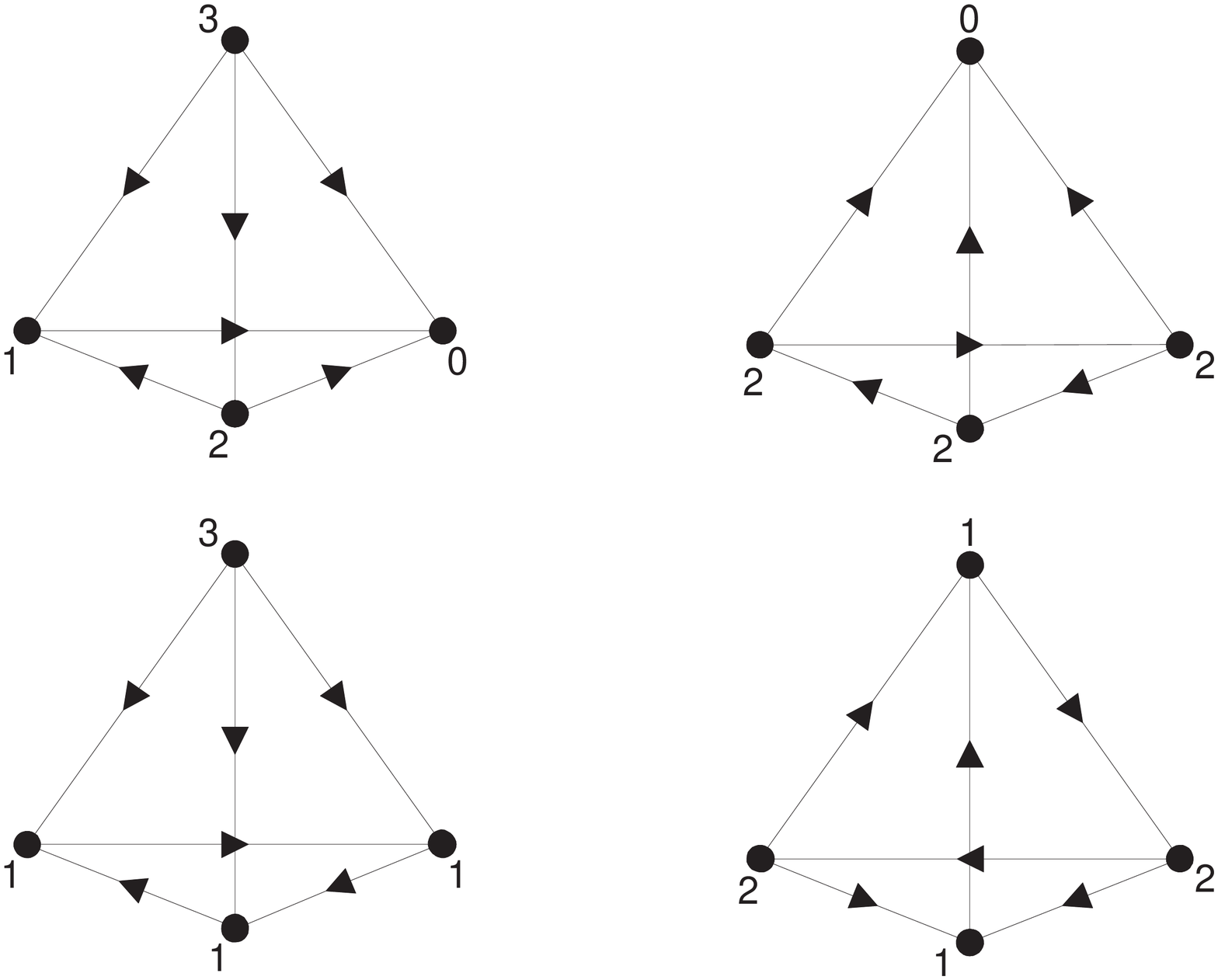';file-properties "XNPEU";}}If we allow for
different commutation factors between different grading groups, then the 
{\em pairs} of groups become distinguishable. On the other hand, with three
points, it is possible to build a symmetric oriented graph. Here are the two
possible oriented graphs with three points:\FRAME{dtbpF}{7.1368cm}{2.8974cm}{%
0pt}{}{\Qlb{trigraph}}{graphtri.eps}{\special{language "Scientific
Word";type "GRAPHIC";maintain-aspect-ratio TRUE;display "PICT";valid_file
"F";width 7.1368cm;height 2.8974cm;depth 0pt;original-width
839.625pt;original-height 347.25pt;cropleft "0";croptop "1";cropright
"1";cropbottom "0";filename 'c: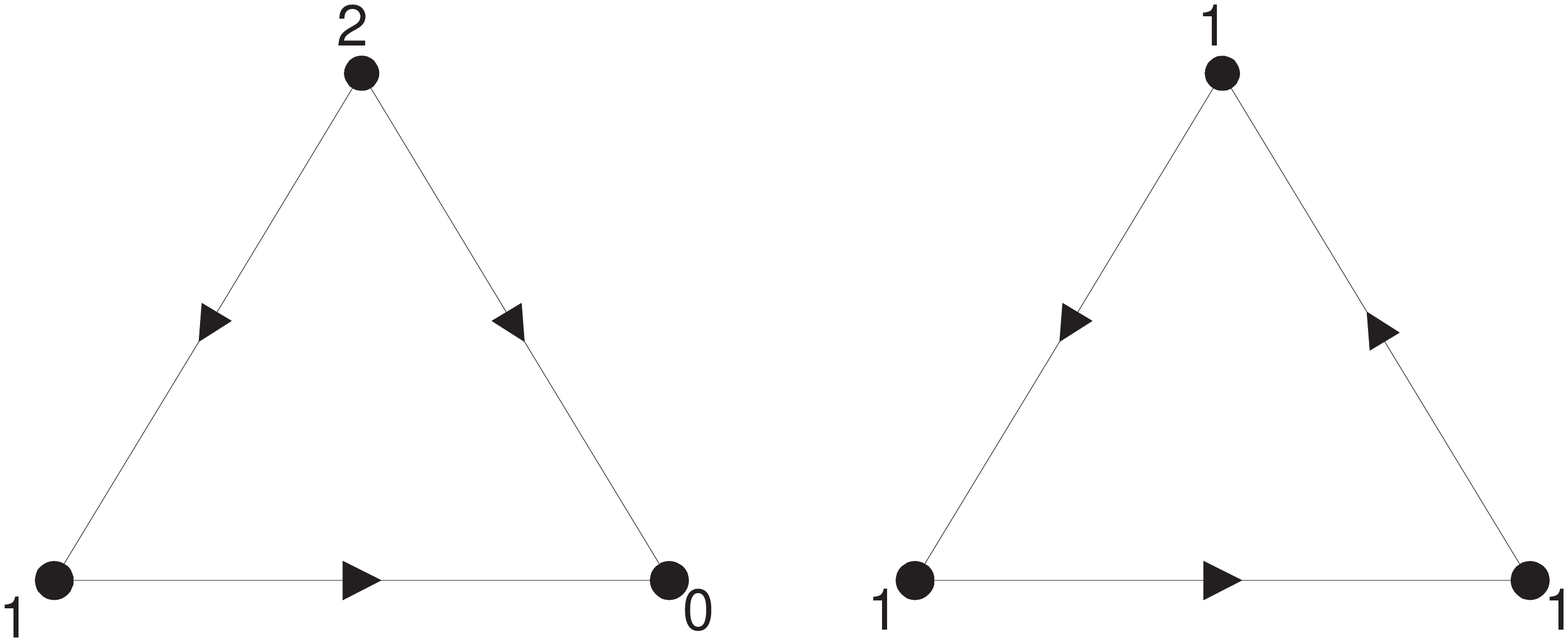';file-properties
"XNPEU";}}Note that the paragrassmann algebras with three and four
generators are described by the first triangle and tetrahedron. The second
triangle is perfectly symmetric in the three points and can be used to build
our generalized Grassmann algebra. We chose here to label each ${\Bbb Z}_{n}$
group with a color, red, green or blue and to note $a_{R}$, $a_{G}$ and $%
a_{B}$ the components in the red, green and blue ${\Bbb Z}_{n}$ groups of an
element $a$ of the grading group. The corresponding commutation factor is 
\begin{eqnarray*}
\epsilon (a,b)
&=&(-1)^{a_{R}b_{R}+a_{G}b_{G}+a_{B}b_{B}}(q)^{a_{R}b_{G}-a_{G}b_{R}+a_{G}b_{B}-a_{B}b_{G}+a_{B}b_{R}-a_{R}b_{B}}=
\\
&=&(-1)^{a_{R}b_{R}+a_{G}b_{G}+a_{B}b_{B}}(q)^{a_{R}b_{G}+a_{G}b_{B}+a_{B}b_{R}}(q^{-1})^{a_{R}b_{B}+a_{B}b_{G}+a_{G}b_{R}}
\end{eqnarray*}
and the smallest grading group is ${\Bbb Z}_{n}^{3}={\Bbb Z}_{n}\times {\Bbb %
Z}_{n}\times {\Bbb Z}_{n}$ if $q$ is an $n^{{\rm th}}$ root of $1$ with $n$
even, ${\Bbb Z}_{2n}^{3}$ if $q$ is an $n^{{\rm th}}$ root of $1$ with $n$
odd, or ${\Bbb Z}^{3}$ if $q$ is not a root of $1$ (in this case, we set $%
n=0 $).

Let us examine in detail the generators one can have in our algebra and
their commutation relations. We can have monochromatic generators having
only one of their three colors equal to one. Then, one can have
``anti-monochromatic'' generators that have one of their colors equal to $-1$%
. These are the conjugates of monochromatic generators. Then, one can have
``black'' generators with all their colors zero and ``white'' generators
with their three colors all equal but different from zero. There are other
combinations but we won't concentrate on them for the moment. Let us denote
by $\theta _{A_{R}}$, $\theta _{A_{G}}$, $\theta _{A_{B}}$ the red, green
and blue monochromatic generators, by $\bar{\theta}_{\bar{A}_{\bar{R}}}$, $%
\bar{\theta}_{\bar{A}_{\bar{G}}}$, $\bar{\theta}_{\bar{A}_{\bar{B}}}$ the
red, green and blue anti-monochromatic generators, by $e_{a}$ the black
generators and by $\eta _{\alpha }$ (resp. $\bar{\eta}_{\dot{\alpha}}$) the
white (resp. anti-white, that is, with all colors equal to $-1$) generators.
We summarize this in the following table: 
\[
\begin{tabular}{c|c|c|c|c|c|c|c|c|c|}
\cline{2-10}
\multicolumn{1}{c|}{} & $\theta _{R}$ & $\theta _{G}$ & $\theta _{B}$ & $%
\bar{\theta}_{\bar{R}}$ & $\bar{\theta}_{\bar{G}}$ & $\bar{\theta}_{\bar{B}}$
& $e$ & $\eta $ & $\bar{\eta}$ \\ \hline
\multicolumn{1}{|c|}{Red} & $1$ & $0$ & $0$ & $-1$ & $0$ & $0$ & $0$ & $1$ & 
$-1$ \\ \hline
\multicolumn{1}{|c|}{Green} & $0$ & $1$ & $0$ & $0$ & $-1$ & $0$ & $0$ & $1$
& $-1$ \\ \hline
\multicolumn{1}{|c|}{Blue} & $0$ & $0$ & $1$ & $0$ & $0$ & $-1$ & $0$ & $1$
& $-1$ \\ \hline
\end{tabular}
\]
and take the convention that capitalized latin indices refer to
monochromatic degrees of freedom, small latin indices to black degrees of
freedom and greek indices to white degrees of freedom.

First, one can note that our wish to have a symmetric fermionic sector is
fulfilled: the black and white sectors form a subalgebra that is an usual
Grassmann algebra: 
\[
\eta _{\alpha }\eta _{\beta }=-\eta _{\beta }\eta _{\alpha }\text{,\quad }%
\bar{\eta}_{\dot{\alpha}}\bar{\eta}_{\dot{\beta}}=-\bar{\eta}_{\dot{\beta}}%
\bar{\eta}_{\dot{\alpha}}\text{,\quad }\eta _{\alpha }\bar{\eta}_{\dot{\beta}%
}=-\bar{\eta}_{\dot{\beta}}\eta _{\alpha } 
\]
\[
e_{a}e_{b}=e_{b}e_{a}\text{,\quad }e_{a}\eta _{\alpha }=\eta _{\alpha }e_{a}%
\text{,\quad }e_{a}\bar{\eta}_{\dot{\alpha}}=\bar{\eta}_{\dot{\alpha}}e_{a} 
\]
An important remark is that the only combinations of monochromatic
generators that can be included in this Grassmann subalgebra are the $\theta
_{R}\theta _{G}\theta _{B}$ (and all similar terms with the three colors in
a different order), the $\bar{\theta}_{\bar{R}}\bar{\theta}_{\bar{G}}\bar{%
\theta}_{\bar{B}}$, the $\theta _{R}\bar{\theta}_{\bar{R}}$, $\theta _{G}%
\bar{\theta}_{\bar{G}}$, $\theta _{B}\bar{\theta}_{\bar{B}}$, the product of 
$n$ monochromatic generators of the same color, and their products. Any
other combination destroys the ${\Bbb Z}_{2}$-grading if the colors are
maintained on an equal footing. That is, if we include $\theta _{R}$, we
must also include $\theta _{G}$ and $\theta _{B}$, which destroys the ${\Bbb %
Z}_{2}$-grading because $\theta _{R}\theta _{G}=q\theta _{G}\theta _{R}$.
These combinations bear a very strong formal resemblance with the only
physically observable combinations of quarks in QCD, especially for $n=0$.
But where QCD gives a dynamical explanation for the confinement, our model
would derive the rules of confinement from algebraical and statistical
arguments: a Fock space generated by operators obeying our commutation rules
would have some states that, because of their colored content, would not be
entirely symmetric or antisymmetric and so would be physically unobservable.
On the other hand, states containing only black or white combinations of
monochromatic generators would have correct symmetry properties with respect
to the exchange of the physically observable particles, that is hadrons.
Please note that we have not imposed any condition on the gauge theory but
only on the grading group. But because of the assumed symmetry between the
three colors, $SU(3)$ and QCD seem very natural in this framework.

The meson-like combinations of monochromatic generators, the $\theta _{R}%
\bar{\theta}_{\bar{R}}$, $\theta _{G}\bar{\theta}_{\bar{G}}$, and $\theta
_{B}\bar{\theta}_{\bar{B}}$, commuting with the rest of the algebra, are 
{\em bosons}. The baryon-like combinations, $\theta _{R}\theta _{G}\theta
_{B}$ and $\bar{\theta}_{\bar{R}}\bar{\theta}_{\bar{G}}\bar{\theta}_{\bar{B}%
} $, anticommuting with each other, with the white generators $\eta $ and $%
\bar{\eta}$ and their odd combinations, are {\em fermions}. Individually,
the monochromatic generators commute with the even sector of the algebra
(that is the bosons) and anticommute with the odd sector (that is the
fermions) and so from our black and white world look like fermions, but to
each other, they do not act like fermions, but rather like parafermions: 
\[
\theta _{R}\theta _{G}=q\theta _{G}\theta _{R}\text{,\quad }\theta
_{G}\theta _{B}=q\theta _{B}\theta _{G}\text{,\quad }\theta _{B}\theta
_{R}=q\theta _{R}\theta _{B} 
\]
\[
\bar{\theta}_{\bar{R}}\bar{\theta}_{\bar{G}}=q\bar{\theta}_{\bar{G}}\bar{%
\theta}_{\bar{R}}\text{,\quad }\bar{\theta}_{\bar{G}}\bar{\theta}_{\bar{B}}=q%
\bar{\theta}_{\bar{B}}\bar{\theta}_{\bar{G}}\text{,\quad }\bar{\theta}_{\bar{%
B}}\bar{\theta}_{\bar{R}}=q\bar{\theta}_{\bar{R}}\bar{\theta}_{\bar{B}} 
\]
with important differences, among which $(\theta _{i})^{2}=\theta _{i}\theta
_{i}=-\theta _{i}\theta _{i}=0$, which indicates that, as fermions, two
monochromatic particles would be unable to be in the same state at the same
point. From these relations, we easily show that the $\theta $'s obey the
ternary rule $\theta _{R}\theta _{G}\theta _{B}=\theta _{G}\theta _{B}\theta
_{R}=\theta _{B}\theta _{R}\theta _{G}$ and not $\theta _{R}\theta
_{G}\theta _{B}=j\theta _{G}\theta _{B}\theta _{R}=j^{2}\theta _{B}\theta
_{R}\theta _{G}$ which was the fundamental relation of the generalized
Grassmann algebras Kerner introduced in \cite{kernergrass}.

\section{Derivations of the generalized Grassmann algebra}

Like one built the supersymmetry generators with the ordinary Grassmann
algebra, one can build with this tricolor Grassmann algebra the generators
of a tricolor supersymmetry. First, let us define the derivation operators
on the algebra of tricolor Grassmann valued functions from the twisted
Leibniz rule: 
\[
\lbrack \partial ^{x},X_{y}]_{c}=\delta _{y}^{x}{\bf 1} 
\]
where, by $x$ and $y$ we mean any kind of index, colored or not, and by $X$,
the multiplication by any generator of the tricolor Grassmann algebra,
colored or not. The colored commutator is defined by 
\[
\lbrack A,B]_{c}=AB-\varepsilon (g_{A},g_{B})BA 
\]
where $g_{A}$ and $g_{B}$ are the grades of $A$ and $B$. The red (resp.
green, blue) derivation operators are attributed the red (resp. green, blue)
grade $-1$. The $\partial ^{a}$ are black, the $\partial ^{\alpha }$ have
all their colors equal to $-1$ and the $\bar{\partial}^{\dot{\alpha}}$ have
all their colors equal to $1$. Thus, the derivations on black and white
indices obey the usual superleibniz rule.

From these rules, we can derive for example 
\[
\partial ^{A_{R}}(\theta _{C_{G}}f)=q^{-1}\theta _{C_{G}}\partial ^{A_{R}}f%
\text{,\quad }\partial ^{A_{R}}(\theta _{C_{R}}\theta _{D_{R}})=\delta
_{C_{R}}^{A_{R}}\theta _{D_{R}}-\delta _{D_{R}}^{A_{R}}\theta _{C_{R}}\text{%
,\quad }\partial ^{A_{R}}(\theta _{A_{R}}f)=f-\theta _{A_{R}}\partial
^{A_{R}}f 
\]
The colors the derivation operators have been assigned are consistent with
the fact that the colored commutator of any pair of derivation operators
vanishes. For example, 
\[
\lbrack \partial ^{R},\partial ^{G}]_{c}=\partial ^{R}\partial
^{G}-q\partial ^{G}\partial ^{R}=0 
\]
In other words, the derivations have the {\em same} commutation relations as
the corresponding generators.

\section{The colored Supersymmetry generators}

\label{susygen}Let us redefine first the usual supersymmetry generators: 
\[
D^{\alpha }=\partial ^{\alpha }+\sigma _{a}^{\alpha \dot{\alpha}}\bar{\eta}_{%
\dot{\alpha}}\partial ^{a}\text{\quad and\quad }\bar{D}^{\dot{\alpha}}=\bar{%
\partial}^{\dot{\alpha}}+\sigma _{a}^{\alpha \dot{\alpha}}\eta _{\alpha
}\partial ^{a} 
\]
These operators are exactly identical to the usual supersymmetry generators
and have the same properties, in particular 
\[
\lbrack D^{\alpha },\bar{D}^{\dot{\alpha}}]_{c}=\{D^{\alpha },\bar{D}^{\dot{%
\alpha}}\}=2\sigma _{a}^{\alpha \dot{\alpha}}\partial ^{a} 
\]
that is, the supersymmetric translations are ``square roots'' of the
translations of the commutative sector. Similar, but richer properties can
be derived for the tricolor supersymmetry generators: 
\[
D^{A_{R}}=\partial ^{A_{R}}+\rho _{\alpha }^{A_{R}C_{G}D_{B}}\theta
_{C_{G}}\theta _{D_{B}}D^{\alpha }+\omega _{a}^{A_{R}\bar{A}_{\bar{R}}}\bar{%
\theta}_{\bar{A}_{\bar{R}}}\partial ^{a} 
\]
\[
D^{C_{G}}=\partial ^{C_{G}}+\rho _{\alpha }^{A_{R}C_{G}D_{B}}\theta
_{D_{B}}\theta _{A_{R}}D^{\alpha }+\omega _{a}^{C_{G}\bar{C}_{\bar{G}}}\bar{%
\theta}_{\bar{C}_{\bar{G}}}\partial ^{a} 
\]
\[
D^{D_{B}}=\partial ^{D_{B}}+\rho _{\alpha }^{A_{R}C_{G}D_{B}}\theta
_{A_{R}}\theta _{C_{G}}D^{\alpha }+\omega _{a}^{D_{B}\bar{D}_{\bar{B}}}\bar{%
\theta}_{\bar{D}_{\bar{B}}}\partial ^{a} 
\]
and their conjugates 
\[
\bar{D}^{\bar{A}_{\bar{R}}}=\bar{\partial}^{\bar{A}_{\bar{R}}}+\bar{\rho}_{%
\dot{\alpha}}^{\bar{A}_{\bar{R}}\bar{C}_{\bar{G}}\bar{D}_{\bar{B}}}\bar{%
\theta}_{\bar{C}_{\bar{G}}}\bar{\theta}_{\bar{D}_{\bar{B}}}\bar{D}^{\dot{%
\alpha}}+\bar{\omega}_{a}^{\bar{A}_{\bar{R}}A_{R}}\theta _{A_{R}}\partial
^{a} 
\]
\[
\bar{D}^{\bar{C}_{\bar{G}}}=\bar{\partial}^{\bar{C}_{\bar{G}}}+\bar{\rho}_{%
\dot{\alpha}}^{\bar{A}_{\bar{R}}\bar{C}_{\bar{G}}\bar{D}_{\bar{B}}}\bar{%
\theta}_{\bar{D}_{\bar{B}}}\bar{\theta}_{\bar{A}_{\bar{R}}}\bar{D}^{\dot{%
\alpha}}+\bar{\omega}_{a}^{\bar{C}_{\bar{G}}C_{G}}\theta _{C_{G}}\partial
^{a} 
\]
\[
\bar{D}^{\bar{D}_{\bar{B}}}=\bar{\partial}^{\bar{D}_{\bar{B}}}+\bar{\rho}_{%
\dot{\alpha}}^{\bar{A}_{\bar{R}}\bar{C}_{\bar{G}}\bar{D}_{\bar{B}}}\bar{%
\theta}_{\bar{A}_{\bar{R}}}\bar{\theta}_{\bar{C}_{\bar{G}}}\bar{D}^{\dot{%
\alpha}}+\bar{\omega}_{a}^{\bar{D}_{\bar{B}}D_{B}}\theta _{D_{B}}\partial
^{a} 
\]
We also define the following alternative operators where the $\theta $'s in
the second term of the definition of the $D$ operators have been swapped: 
\[
D^{\prime A_{R}}=\partial ^{A_{R}}+\rho _{\alpha }^{A_{R}C_{G}D_{B}}\theta
_{D_{B}}\theta _{C_{G}}D^{\alpha }+\omega _{a}^{A_{R}\bar{A}_{\bar{R}}}\bar{%
\theta}_{\bar{A}_{\bar{R}}}\partial ^{a} 
\]
$D^{\prime C_{G}}$, $D^{\prime D_{B}}$, $\bar{D}^{\prime \bar{A}_{\bar{R}}}$%
, $\bar{D}^{\prime \bar{C}_{\bar{G}}}$ and $\bar{D}^{\prime \bar{D}_{\bar{B}%
}}$ are defined similarly.

It is straightforward to show that
\begin{eqnarray*}
\lbrack D^{A_{R}},[D^{\prime A_{G}},D^{A_{B}}]_{c}]_{c}
&=&[D^{A_{G}},[D^{\prime A_{B}},D^{A_{R}}]_{c}]_{c}=[D^{A_{B}},[D^{\prime
A_{R}},D^{A_{G}}]_{c}]_{c}= \\
&=&(q-1)\rho _{\alpha }^{A_{R}A_{G}A_{B}}D^{\alpha }
\end{eqnarray*}
and similar relations with barred operators. That is, the monochromatic
supersymmetric translations are also {\em cubic} roots of the ordinary {\em %
supersymmetric} translations.

\section{The ${\Bbb Z}_{n}^{3}$-graded $\varepsilon $-Lie superalgebra}

The algebra generated by the $D$, $\bar{D}$ and $\partial $ operators by
means of the colored commutator is finite-dimensional (if the generalized
Grassmann algebra is generated by a finite number of generators, of course).
In fact, our colored supersymmetric translations generate an $\varepsilon $%
-Lie superalgebra\cite{sheunert}, where the product is the colored
commutator.

To write the commutation relations, we need to define the other elements of
the algebra: 
\[
Q^{A_{R}A_{G}}=\rho _{\beta }^{A_{R}A_{G}B_{B}}\theta _{B_{B}}D^{\beta }%
\text{,\quad }Q^{A_{R}\dot{\alpha}}=\rho _{\beta }^{A_{R}B_{G}B_{B}}\sigma
_{b}^{\beta \dot{\alpha}}\theta _{B_{G}}\theta _{B_{B}}\partial ^{b}\text{%
,\quad }Q^{A_{R}A_{G}\dot{\alpha}}=\rho _{\beta }^{A_{R}A_{G}B_{B}}\sigma
_{b}^{\beta \dot{\alpha}}\theta _{B_{B}}\partial ^{b}
\]
\[
Q^{A_{R}A_{G}\bar{A}_{\bar{R}}\bar{A}_{\bar{B}}}=\rho _{\beta
}^{A_{R}A_{G}B_{B}}\bar{\rho}_{\dot{\beta}}^{\bar{A}_{\bar{R}}\bar{B}_{\bar{G%
}}\bar{A}_{\bar{B}}}\sigma _{b}^{\beta \dot{\beta}}\theta _{B_{B}}\bar{\theta%
}_{\bar{B}_{\bar{G}}}\partial ^{b}\text{,\quad }Q^{A_{R}A_{G}\bar{A}_{\bar{R}%
}}=\rho _{\beta }^{A_{R}A_{G}B_{B}}\bar{\rho}_{\dot{\beta}}^{\bar{A}_{\bar{R}%
}\bar{B}_{\bar{G}}\bar{B}_{\bar{B}}}\sigma _{b}^{\beta \dot{\beta}}\theta
_{B_{B}}\bar{\theta}_{\bar{B}_{\bar{G}}}\bar{\theta}_{\bar{B}_{\bar{B}%
}}\partial ^{b}
\]
\[
Q^{A_{R}\bar{A}_{\bar{G}}}=\rho _{\beta }^{A_{R}B_{G}B_{B}}\bar{\rho}_{\dot{%
\beta}}^{\bar{B}_{\bar{R}}\bar{A}_{\bar{G}}\bar{B}_{\bar{B}}}\sigma
_{b}^{\beta \dot{\beta}}\theta _{B_{G}}\theta _{B_{B}}\bar{\theta}_{\bar{B}_{%
\bar{B}}}\bar{\theta}_{\bar{B}_{\bar{R}}}\partial ^{b}
\]
and the similar operators in other colors, or conjugated. The commutation
relations that haven't been given in the previous section are 
\begin{eqnarray*}
\lbrack D^{A_{R}},\bar{D}^{\bar{A}_{\bar{R}}}]_{c} &=&q[D^{\prime A_{R}},%
\bar{D}^{\bar{A}_{\bar{R}}}]_{c}=q[D^{A_{R}},\bar{D}^{\prime \bar{A}_{\bar{R}%
}}]_{c}=q^{2}[D^{\prime A_{R}},\bar{D}^{\prime \bar{A}_{\bar{R}}}]_{c}= \\
&=&Q^{A_{R}\bar{A}_{\bar{R}}}+(\omega _{a}^{A_{R}\bar{A}_{\bar{R}}}+\bar{%
\omega}_{a}^{\bar{A}_{\bar{R}}A_{R}})\partial ^{a}
\end{eqnarray*}
\[
\lbrack D^{A_{R}},\bar{D}^{\bar{A}_{\bar{G}}}]_{c}=q[D^{\prime A_{R}},\bar{D}%
^{\bar{A}_{\bar{G}}}]_{c}=q[D^{A_{R}},\bar{D}^{\prime \bar{A}_{\bar{G}%
}}]_{c}=q^{2}[D^{\prime A_{R}},\bar{D}^{\prime \bar{A}_{\bar{G}%
}}]_{c}=Q^{A_{R}\bar{A}_{\bar{G}}}
\]
\[
\lbrack D^{A_{R}},\bar{D}^{\bar{A}_{\bar{B}}}]_{c}=q[D^{\prime A_{R}},\bar{D}%
^{\bar{A}_{\bar{B}}}]_{c}=q[D^{A_{R}},\bar{D}^{\prime \bar{A}_{\bar{B}%
}}]_{c}=q^{2}[D^{\prime A_{R}},\bar{D}^{\prime \bar{A}_{\bar{B}%
}}]_{c}=Q^{A_{R}\bar{A}_{\bar{B}}}
\]
\[
\lbrack D^{A_{B}},D^{\prime A_{R}}]_{c}=-[D^{\prime
A_{B}},D^{A_{R}}]_{c}=(1-q)Q^{A_{B}A_{R}}
\]
\[
\lbrack D^{A_{R}},\bar{D}^{\dot{\alpha}}]_{c}=q[D^{\prime A_{R}},\bar{D}^{%
\dot{\alpha}}]_{c}=Q^{A_{R}\dot{\alpha}}
\]
\[
\lbrack Q^{AB},\bar{Q}^{\bar{A}\bar{B}}]=-Q^{AB\bar{A}\bar{B}}\text{,\quad }[%
D^{\dot{\alpha}},Q^{AB}]=-Q^{AB\dot{\alpha}}
\]
\[
\lbrack D^{A_{R}},Q^{A_{G}A_{B}}]_{c}=[D^{\prime
A_{R}},Q^{A_{G}A_{B}}]_{c}=\rho _{\beta }^{A_{R}A_{G}A_{B}}D^{\beta }
\]
\[
\lbrack D^{A_{R}},Q^{A_{G}\dot{\alpha}}]_{c}=q[D^{A_{G}},Q^{A_{R}\dot{\alpha}%
}]_{c}=qQ^{A_{R}A_{G}\dot{\alpha}}
\]
\[
\lbrack D^{A_{R}},Q^{A_{G}A_{B}\dot{\alpha}}]_{c}=[D^{\prime
A_{R}},Q^{A_{G}A_{B}\dot{\alpha}}]_{c}=\rho _{\beta
}^{A_{R}A_{G}A_{B}}\sigma _{b}^{\beta \dot{\alpha}}\partial ^{b}
\]
\[
\lbrack D^{A_{R}},Q^{A_{G}A_{B}\bar{A}\bar{B}}]_{c}=[D^{\prime
A_{R}},Q^{A_{G}A_{B}\bar{A}\bar{B}}]_{c}=\rho _{\beta }^{A_{R}A_{G}B_{B}}Q^{%
\bar{A}\bar{B}\beta }
\]
\[
\lbrack \bar{D}^{\bar{A}_{\bar{R}}},Q^{A_{R}A_{G}\bar{A}_{\bar{G}}\bar{A}_{%
\bar{B}}}]_{c}=[\bar{D}^{\prime \bar{A}_{\bar{R}}},Q^{A_{R}A_{G}\bar{A}_{%
\bar{G}}\bar{A}_{\bar{B}}}]_{c}=q^{-1}\bar{\rho}_{\dot{\beta}}^{\bar{A}_{%
\bar{R}}\bar{A}_{\bar{G}}\bar{A}_{\bar{B}}}Q^{A_{R}A_{G}\dot{\beta}}
\]
\[
\lbrack \bar{D}^{\bar{A}_{\bar{R}}},Q^{A_{G}A_{B}\bar{A}_{\bar{G}}\bar{A}_{%
\bar{B}}}]_{c}=[\bar{D}^{\prime \bar{A}_{\bar{R}}},Q^{A_{G}A_{B}\bar{A}_{%
\bar{G}}\bar{A}_{\bar{B}}}]_{c}=-\bar{\rho}_{\dot{\beta}}^{\bar{A}_{\bar{R}}%
\bar{A}_{\bar{G}}\bar{A}_{\bar{B}}}Q^{A_{G}A_{B}\dot{\beta}}
\]
\[
\lbrack \bar{D}^{\bar{A}_{\bar{R}}},Q^{A_{B}A_{R}\bar{A}_{\bar{G}}\bar{A}_{%
\bar{B}}}]_{c}=[\bar{D}^{\prime \bar{A}_{\bar{R}}},Q^{A_{B}A_{R}\bar{A}_{%
\bar{G}}\bar{A}_{\bar{B}}}]_{c}=q\bar{\rho}_{\dot{\beta}}^{\bar{A}_{\bar{R}}%
\bar{A}_{\bar{G}}\bar{A}_{\bar{B}}}Q^{A_{B}A_{R}\dot{\beta}}
\]
\[
\lbrack D^{A_{R}},Q^{A_{G}A_{B}\bar{A}}]_{c}=[D^{\prime A_{R}},Q^{A_{G}A_{B}%
\bar{A}}]_{c}=\rho _{\beta }^{A_{R}A_{G}B_{B}}Q^{\bar{A}\beta }
\]
\[
\lbrack \bar{D}^{\bar{A}_{\bar{R}}},Q^{A_{R}A_{G}\bar{A}_{\bar{G}}}]_{c}=[%
\bar{D}^{\prime \bar{A}_{\bar{R}}},Q^{A_{R}A_{G}\bar{A}_{\bar{G}%
}}]_{c}=Q^{A_{R}A_{G}\bar{A}_{\bar{R}}\bar{A}_{\bar{G}}}
\]
\[
\lbrack \bar{D}^{\bar{A}_{\bar{R}}},Q^{A_{G}A_{B}\bar{A}_{\bar{G}}}]_{c}=[%
\bar{D}^{\prime \bar{A}_{\bar{R}}},Q^{A_{G}A_{B}\bar{A}_{\bar{G}%
}}]_{c}=-qQ^{A_{G}A_{B}\bar{A}_{\bar{R}}\bar{A}_{\bar{G}}}
\]
\[
\lbrack \bar{D}^{\bar{A}_{\bar{R}}},Q^{A_{B}A_{R}\bar{A}_{\bar{G}}}]_{c}=[%
\bar{D}^{\prime \bar{A}_{\bar{R}}},Q^{A_{B}A_{R}\bar{A}_{\bar{G}%
}}]_{c}=q^{2}Q^{A_{B}A_{R}\bar{A}_{\bar{R}}\bar{A}_{\bar{G}}}
\]
\[
\lbrack D^{A_{R}},Q^{A_{G}\bar{A}}]_{c}=[D^{\prime A_{R}},Q^{A_{G}\bar{A}%
}]_{c}=qQ^{A_{R}A_{G}\bar{A}}
\]
\[
\lbrack \bar{D}^{\bar{A}_{\bar{R}}},Q^{A_{R}\bar{A}_{\bar{B}}}]_{c}=[\bar{D}%
^{\prime \bar{A}_{\bar{R}}},Q^{A_{R}\bar{A}_{\bar{B}}}]_{c}=Q^{A_{R}\bar{A}_{%
\bar{B}}\bar{A}_{\bar{R}}}
\]
\[
\lbrack \bar{D}^{\bar{A}_{\bar{R}}},Q^{A_{G}\bar{A}_{\bar{B}}}]_{c}=[\bar{D}%
^{\prime \bar{A}_{\bar{R}}},Q^{A_{G}\bar{A}_{\bar{B}}}]_{c}=-q^{-1}Q^{A_{G}%
\bar{A}_{\bar{B}}\bar{A}_{\bar{R}}}
\]
\[
\lbrack \bar{D}^{\bar{A}_{\bar{R}}},Q^{A_{B}\bar{A}_{\bar{B}}}]_{c}=[\bar{D}%
^{\prime \bar{A}_{\bar{R}}},Q^{A_{B}\bar{A}_{\bar{B}}}]_{c}=-qQ^{A_{B}\bar{A}%
_{\bar{B}}\bar{A}_{\bar{R}}}
\]
and of course all the similar relations obtained by simultaneously replacing 
$R$ (resp. $G$, $B$, $\bar{R}$, $\bar{G}$, $\bar{B}$) by $G$ (resp. $B$, $R$%
, $\bar{G}$, $\bar{B}$, $\bar{R}$) or by $B$ (resp. $R$, $G$, $\bar{B}$, $%
\bar{R}$, $\bar{G}$) and/or by simultaneously conjugating all operators, $%
\rho $ and $\bar{\rho}$. All other commutation relations give zero.

The generalized Jacobi identity is {\em always} satisfied: 
\[
\varepsilon (g_{A},g_{C})[[A,B]_{c},C]_{c}+\varepsilon
(g_{B},g_{A})[[B,C]_{c},A]_{c}+\varepsilon (g_{C},g_{B})[[C,A]_{c},B]_{c}=0 
\]
for any $A$, $B$ and $C$ in the algebra. The main consequence of this is
that it is isomorphic to an ordinary Lie superalgebra\cite{sheunert}.

\end{document}